\documentclass[]{mn2e}
\usepackage{graphicx}

\newif\ifAMStwofonts
\def\gs{\mathrel{\raise0.35ex\hbox{$\scriptstyle >$}\kern-0.6em 
\lower0.40ex\hbox{$\scriptstyle \sim$}}}
\def\ls{\mathrel{\raise0.35ex\hbox{$\scriptstyle <$}\kern-0.6em 
\lower0.40ex\hbox{$\scriptstyle \sim$}}}

\newcommand{\be}{\begin{equation}}
\newcommand{\ee}{\end{equation}}
\newcommand{\bea}{\begin{eqnarray}}
\newcommand{\eea}{\end{eqnarray}}

\ifCUPmtlplainloaded \else
  \ifAMStwofonts \else % If no AMS fonts
    \def\upi{\pi}
                 \def\umu{\mu}
    \def\upartial{\partial}
  \fi
\fi

\title{The redshift distribution of gamma-ray bursts revisited}
\author[Natarajan et al.]{P. Natarajan$^{1,2}$, B. Albanna$^{2,3}$, 
J. Hjorth$^4$, E. Ramirez-Ruiz$^5$, N. Tanvir$^6$ \& R. Wijers$^7$\\
$^1$ Astronomy Department, Yale University, P.O. Box 208101, 
       New Haven, CT 06520-8101, USA \\
$^2$Department of Physics, Yale University, P. O. Box 208101,
New Haven, CT 06520-8101, USA\\
$^3$Department of Physics, Univ. of California at Berkeley, 
366 Le Conte Hall, Berkeley, CA 94720-7300, USA \\ 
$^4$Niels Bohr Institute,
University of Copenhagen, DK-2100 Copenhagen, DENMARK\\
$^5$Institute for Advanced Study, Princeton, NJ, USA\\
$^6$ Centre for Astrophysics Research, University of Hertfordshire,
College Lane, Hatfield AL10 9AB, UK\\
$^7$Astronomical Institute, Faculty of Science, University of
Amsterdam, Kruislaan 403, 1098 SJ Amsterdam, The Netherlands}

\pagerange{\pageref{firstpage}--\pageref{lastpage}}
\pubyear{1994}

\begin{document}

\maketitle

\label{firstpage}

\begin{abstract}
In this letter, we calculate the redshift distribution of gamma-ray
bursts assuming that they trace (i) the globally averaged star
formation rate or (ii) the average metallicity in the Universe. While
at redshifts 5 and below, both the star formation rate and metallicity
are observationally determined modulo some uncertainties, at higher
redshifts there are few constraints. We extrapolate the star formation
rate and metallicity to higher redshifts and explore models that are
broadly consistent with bounds on the optical depth from WMAP
results. In addition, we also include parametric descriptions of the
luminosity function, and the typical spectrum for GRBs.  With these
essential ingredients included in the modeling, we find that a
substantial fraction 75\% of GRBs are expected to originate at
redshifts below 4, in variance with some previous estimates.
Conversely, if we assume as expected for the collapsar model that
gamma-ray bursts favour a low metallicity environment and therefore,
relate the GRB rate to a simple model of the average metallicity as a
function of redshift, we find that a higher fraction of bursts, about
40\% originate from $z > 4$. We conclude with the implications of
SWIFT GRB detections.
\end{abstract}


\section{Introduction}

The demonstration that long-duration gamma-ray bursts (GRBs) are
related to core-collapse supernovae (Galama et al. 1998; Stanek et
al. 2003; Hjorth et al. 2003; Malesani et al. 2004), likely leading to
the formation of black holes in `collapsars' (MacFadyen \& Woosley
1999) suggests, that GRBs trace the deaths (and hence births) of
short-lived massive stars. Moreover, as GRBs can be detected to very
high redshifts (Lamb \& Reichart 2000), unhindered by intervening dust
-- the current record is $z= 4.50$ (Andersen et al. 2000) -- they hold
the promise of being useful tracers of star formation in the Universe
(Wijers et al. 1998, Totani 1997, Blain et al. 1999; Blain \&
Natarajan 2000; Ramirez-Ruiz, Trentham \& Blain 2002; Bromm \& Loeb
2002; Gou et al. 2004). This ansatz, that GRBs are likely to
effectively trace the observed star formation rate (SFR), has been
used to predict the redshift distribution of GRBs, despite our lack of
knowledge of SFRs at $z > 6$. Observational estimates of the SFR even
at modest redshifts have been plagued by uncertainties arising due to
correction for dust extinction. Therefore, SFR's need to be
extrapolated to higher redshifts. The only current constraint that is
useful is the WMAP (Wilkinson Microwave Anisotropy Probe) estimate of
the optical depth and the redshift of re-ionization (Kogut et al.
2003), both of which suggest the existence of ionizing sources out to
very high redshifts. The extrapolations of the SFR explored here would
provide the requisite number of ionizing photons as demonstrated by
Somerville \& Livio (2003).

The discovery of several $z \sim 6$ quasars in the SDSS (Sloan Digital
Sky Survey) whose spectra are consistent with showing zero flux below
Lyman-$\alpha$ (a `Gunn-Peterson' trough) indicates that the IGM
(Inter-Galactic Medium) had a significant neutral fraction at $z > 6$
(Fan et al. 2001; Becker et al. 2001). The ionization history of the
Universe has also been constrained via observations of the cosmic
microwave background (CMB). The first-year results from the WMAP
satellite constrain the optical depth to Thomson scattering to be
$\tau = 0.17 \pm 0.04$, implying a re-ionization redshift $z_{\rm
reion} = 17\pm5$ (Kogut et al. 2003). Our extrapolation of the SFR to
higher redshifts is in consonance with these observations.

In this work, we explore a fully self-consistent approach to predict
the expected redshift distribution of GRBs at $z > 3$. In section 2,
the observed redshift distribution of GRBs is presented. A clutch of
star formation models are studied here, which are then extrapolated to
higher redshifts. In section 3, we also investigate the proposition
that GRBs progenitors might be preferentially metal-poor as expected
in the collapsar model and as suggested by the observations of Fynbo
et al. (2003) \& Vreeswijk et al. (2004). A model where the GRB rate,
is inversely correlated to the mean metallicity in the Universe is
explored. GRBs are then modeled with a typical spectral shape and a
luminosity function, the details of which are presented in section
4. We conclude with a synopsis and discussion of our results in the
context of GRB detections by the SWIFT satellite in section 5.

\section{GRBs: their observed redshift distribution and the star
  formation rate}

Despite the ability of spacecraft equipped with GRB detectors to
detect GRBs to high redshift (Gorosabel et al. 2004), no very high-z
GRB has yet been detected at say, $z > 5$. It is important to note
that selection effects are difficult to quantify so the observed
distribution may not be the same as the true distribution. Two of the
primary causes of selection effects are the lack of knowledge of the
intrinsic luminosity function of GRBs and the details of the central
engine that drives the bursts both of which impact the redshift
distribution of bursts.

The recent launch of Swift promises to detect GRBs en masse (Gehrels
et al. 2004). It is interesting to note that $z_{\rm median} = 1.1$
for non-SWIFT bursts, and $z_{\rm median} = 2.9$ for SWIFT bursts,
bolstering hopes that SWIFT may indeed push detection of GRBs more
efficiently to higher redshifts.  While the calibration of the
detection efficiency for SWIFT will be best determined over the next
couple of years of operation, for the purposes of this paper we use
the detection efficiency model curve for SWIFT adopted by several
other authors (Gou et al. 2004; Gorosabel et al. 2004 (Fig.~3);
Porciani \& Madau 2001). The detection efficiency curve as a function
of redshift that we adopt is overplotted in Fig.~2 (thin solid line). 

To what extent do GRBs trace star formation? It has been argued that
individual GRBs may trace galaxies or regions of galaxies with high
specific star formation (Christensen et al. 2004; Courty et al.2004)
or low metallicity (Fynbo et al. 2003; MacFadyen \& Woosley 1999;
Ramirez-Ruiz et al. 2002). However, this does not preclude the
possibility that GRBs trace the star-formation rate of the Universe in
a globally averaged sense. Indeed, the luminosity function (LF) of
$z>2$ GRB host galaxies, assuming GRBs trace UV light, and the LF of
Lyman-break galaxies are consistent (Jakobsson et al. 2005). We start
with the premise that the GRB rate \footnote{Throughout this paper GRB
rate refers to the GRB occurence rate and not the detected rate unless
explicitly stated.} traces the global SFR of the
Universe, $R_{\rm GRB}(z)\propto R_{\rm SF}(z)$, where $R_{\rm SF}(z)$
is the co-moving rate density of star formation.

The expected evolution of the globally averaged cosmic SFR with
redshift has been studied by many authors, following the first
successful attempt by Madau et al. (1996), who based their estimates
on the observed (rest--frame) UV luminosity density of galaxy
populations. Using various observational techniques, the cosmic SFR
can now be traced to $z\approx 5$, although there is no clear
consensus on the details of dust correction both at high and low
redshifts (Dickinson et al. 2003; Steidel et al. 2004) In this paper,
we explore several models that describe (all shown in Fig.~1) the
global SFR per unit co-moving volume. Wherever needed, values for
cosmological parameters consistent with the WMAP results (Spergel et
al. 2003) are assumed: matter density $\Omega_m = 0.3$, baryon density
$\Omega_b =0.044$, dark energy $\Omega_{\Lambda}=0.70$, Hubble
parameter $H_0=70$ km/s/Mpc, fluctuation amplitude $\sigma_8 = 0.9$,
and a scale-free primordial power spectrum $n_s=1$.

The first model studied here (Model I hereafter) is similar to one
considered previously by Porciani \& Madau (2001) which they used in
modeling the fraction of lensed GRBs (their model SF3). Our Model I
(dotted curve in Fig.~1) provides a good fit to the sub-mm
determinations of the luminosity density.  Blain et al. (1999) have
argued that the SFR at all redshifts may have been severely
underestimated due to large amounts of dust extinction detected in
SCUBA galaxies. In addition, we construct a high redshift
extrapolation for a SF history (Model II, the solid curve
in Fig.~1) that is required to fit the observational data at low
redshifts and has sufficient star formation at high redshifts ($z >
10$) to match the WMAP constraints on the optical depth. We justify
this extrapolation with a physical picture in mind using a
semi-analytic model, the details of which are described in the
following section. Models I and II predict very similar GRB rates
although they appear to be divergent at $z > 7$. Finally, as GRBs
themselves do not seem to be pointing to large amounts of dust in
their host galaxies (Berger et al. 2003; Tanvir et al. 2004; Fynbo et
al. 2003; Jakobsson et al. 2005; Vreeswijk et al. 2004), and while
this might be a selection effect, we have also considered a model in
which the bulk of the star formation is not obscured by dust at $z >
5$ but occurs in a population of numerous very faint galaxies that
each may have moderate amounts of dust (our Model III). We have used
the lower limit on the SFR from observations of Lyman-break galaxies
at $z = 5$ to constrain our Model III (dashed curve in Fig.~1).

\subsection{Constructing Model II: a semi-analytic model for high
  redshift star formation}

We calculate the global SFR density from $z \sim 30$--3 using a simple
model that combines the rate of dark matter halo growth with a
prescription for cooling and star formation and match this rate to
observational constraints on the SFR obtained at $3 \la z \la 6$ and
at lower redshifts (similar to previous work by Somerville \& Livio
2003). Prompted by the WMAP estimate of the optical depth
at re-ionization, which points to the existence of a significant
number of ionizing sources at high redshift (assumed to be stars), we
construct a SF history with vigorous activity at the earliest
epochs. This is done in the context of the standard structure
formation scenario within the cold dark matter paradigm, where halos
build-up hierarchically and galaxies form from the condensation of
baryons in dark halos. A much more sophisticated version of this
approach was pioneered (Kauffmann et al.  1993; Cole et al. 1994),
developed and honed over the years by several groups. 

\begin{figure}
\begin{center}
\includegraphics[height=7.5cm,width=7.5cm]{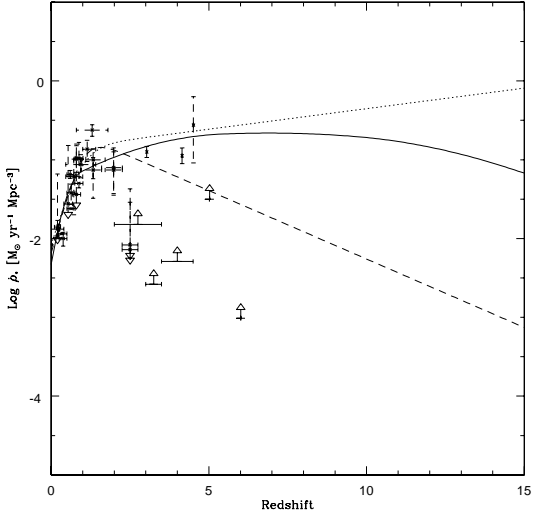}
\caption{The three star formation models considered in this work
  Model I (dotted curve); Model II the constructed 
semi-analytic model (solid curve) and the conservative Model III
  (short dashed curve). The data points (with error bars) are collated from 
the literature of measured star formation rates from various authors,
  this particular compilation was taken  from a paper by Dickinson et
  al. (2003) and references therein.}
\end{center}
\end{figure}

With the abundance of dark matter halo masses $n(M,z)$ determined
using the Press-Schechter formalism, we then proceed to use a simple
cooling criterion to determine the fraction of gas that is converted
into stars modulo some efficiency factor $\epsilon_*$ \footnote{The
requirement to match up with the measured value of the SF at $z = 0$
constrains the value of $\epsilon_*$ to $\sim$ 10\%.} taken to be
roughly 10\% in these collapsed halos. Depending on the primary
coolant, atomic or molecular, there is a critical mass threshold for a
dark matter halo's gas content to cool, and form stars. The star
formation rate can then be written as follows:
\begin{equation}
\dot{\rho}_* = \epsilon_* \rho_b \, \frac{{\rm d}f_m}{\rm
dt}(M>M_{\rm thres}),
\end{equation} 
where $f_m$ is the fraction of the total mass in collapsed halos with
masses greater than $M_{\rm thres}$, obtained from the halo mass
function $dn(M,z)/dM$, $\rho_b$ is the
mean density of baryons and the efficiency of
converting gas into stars is encapsulated in $\epsilon_*$. The threshold 
mass $M_{\rm thres}$ determines the dominant cooling route, it
corresponds to halos with a virial temperature of about $10^4$ K for 
atomic cooling, and  $T \simeq 100$ K for molecular
cooling. Using standard cooling arguments and assuming a Salpeter IMF
for the stars formed, the SF rate can be computed. The predicted SFR for this 
model (solid curve in Fig.~1) is then
calibrated with observational estimates at `low' redshift
$3 \la z \la 6$. 
Using  Somerville \& Livio's estimates for the fraction of ionizing photons available
per hydrogen atom given the SFR, we argue that these SFR models 
are consistent with the optical depth measured by WMAP (for further
details see Section 4.2 of Somerville \& Livio 2003). 

\subsection{GRB rate and metallicity: exploring a toy model}

A larger proporation of the higher redshift GRB host galaxies are
detected as Lyman-$\alpha$ emitters (Kulkarni et al. 1998; Ahn 2000;
M{\o}ller et al. 2002; Vreeswijk et al. 2004; Fynbo et al. 2003)
compared to galaxies selected by the Lyman-break technique (Shapley et
al. 2003) at similar redshifts. This led Fynbo et al. (2003) to
suggest a preference for GRB progenitors to be metal-poor as predicted
by the collapsar model. In the collapsar model, the presence of a
strong stellar wind (a consequence of high metallicity) would hinder
the production of a GRB, therefore metal poor hosts would be favoured
sites (MacFadyen \& Woosley 1999; Izzard et al. 2004). Here, we
explore a toy model wherein the GRB rate decreases with increasing
metallicity. There are observational constraints on the mean
metallicity of the Universe as a function of redshift (Pettini 2003),
however, for our exploratory purposes it is adequate to consider a
simple model (Model IV) wherein the GRB rate is modeled just as a step
function with higher rate at large redshifts ($z \gs 3$) when the
average metallicity of the Universe is low, and taken to have a lower
rate when the Universe is metal-rich at lower redshifts. This
transition in the assumed GRB rate which is assumed to mimic the
change in the mean metallicity of the Universe is taken to occur
abruptly at $z = 3$. While metals are produced as a consequence of SF,
by construction, Model IV bears no relation to a SFR, this was done
for simplicity. The predictions for the expected redshift distribution
of GRBs under these assumptions for Model IV are also plotted in
Fig.~2.
   
\section{Observations of GRBs}

In order to predict the redshift distribution of GRBs, in addition to
a phenomenological model for the SF, we need to model the observed
properties of the bursts, their number counts and luminosity
distribution. The isotropically emitted photon flux $P$ detected
within an energy band $E_{1}<E<E_{2}$ arising from a GRB at redshift
$z$ with a luminosity distance $d_L(z)$ is given by,
\begin{equation}
P={\frac{(1+z) \int_{(1+z)E_{1}}^{(1+z)E_{2}}S(E) \,dE}
{4 \pi d^2_{\rm L}(z)}}\,\,{\rm erg\,s^{-1}},
\end{equation}
where $S(E)$ is the differential rest--frame photon luminosity of the 
source. The total burst luminosity in a given band can then be computed by
integrating $E\,S(E)$ over the relevant energy range. Given a
normalized LF $\psi(L)$ for GRBs, the  burst rate of observed peak
fluxes in the interval $(P_1,P_2)$ is 
\begin{eqnarray}
{dN \over dt}(P_1 \leq P < P_2) = \int_0^\infty
 dz \,\frac{dV(z)}{dz} \frac{R_{\rm GRB}(z)}{1+z} \\
\nonumber \\ \int_{L(P_1,z)}^{L(P_2,z)} dL'\, \psi(L') \epsilon(P)\;,
\label{counts1}
\end{eqnarray}
where $dV/dz$ is the co-moving volume element, $R_{\rm GRB}(z)$ is the 
co-moving GRB rate density, $\epsilon(P)$ is the detector efficiency, 
and the $(1+z)^{-1}$ is the cosmological time dilation factor.
The comoving volume element is given by:
\be
{dV \over dz}= {c\over H_0}\frac{\Delta\omega_s d_L^2(z)}
{(1+z)^2 [\Omega_M (1+z)^3+\Omega_K(1+z)^2+\Omega_\Lambda]^{1/2}},
\ee
where $\Delta\omega_s$ is the solid angle spanned by the survey.
We adopted the detector efficiency function for BATSE provided by
Band et al. (2003) to find our best-fit parameters for the GRB LF. 

GRBs have a broad LF when uncorrected for beaming effects, however the
data are insufficient at the present time to directly determine 
$\psi(L)$ from observations. Therefore, we model the number counts as
done by several authors (Porciani \& Madau 2001; Guetta et al. 2005)
by assuming that the burst luminosity distribution does not evolve
with redshift. A simple parametric form is chosen for $\psi(L)$, 
\be
\psi(L) \propto \, \left(\frac{L}{L_0}\right)^{\gamma}
\exp\left(-\frac{L}{L_0}\right),
\label{LF}
\ee
where $L$ denotes the peak luminosity in the 30--2000 keV energy range 
(rest--frame), $\gamma$ is the asymptotic slope at the bright end, and
$L_0$ is a characteristic cutoff luminosity. The normalization $\int_0^\infty \psi(L)
dL=1$, is used to define the constant of proportionality in
eqn.(6). To describe the typical spectrum of a GRB, we use the form 
proposed by Band et al. (1993):
\begin{eqnarray}
S(E)=A\times  
\left\{ \begin{array}{ll}
%\begin{cases}
\displaystyle{
\left(\frac{E}{100\,{\rm keV}}\right)^\alpha\exp\left[\frac{E
(\beta-\alpha)}{E_b}\right]}\,{E<E_b} \\
\displaystyle{
\left(\frac{E_b}{100\,{\rm keV}}\right)^{\alpha-\beta} \nonumber
\exp{(\beta-\alpha)} 
\left(\frac{E}{100 \,{\rm keV}}\right)^\beta}\,{E\geq E_b}\;
%\end{cases}
\end{array} \right.
\end{eqnarray}
The energy spectral indices, $\alpha$ and $\beta$, have the values
$-1$ and $-2.25$, respectively, measured from the bright BATSE bursts 
by Preece et al. (2000), with a spectral break at $E_b=511$ keV. 
This description has been successfully calibrated against the observed
number counts by Porciani \& Madau (2001) which we adopt. They in turn
used the off--line BATSE sample of Kommers et al. (2000), which includes 1998  
archival BATSE (``triggered'' plus ``non--triggered'') bursts in the 
50--300 keV band. 

\begin{figure}
\begin{center}
\includegraphics[height=9cm,width=9cm]{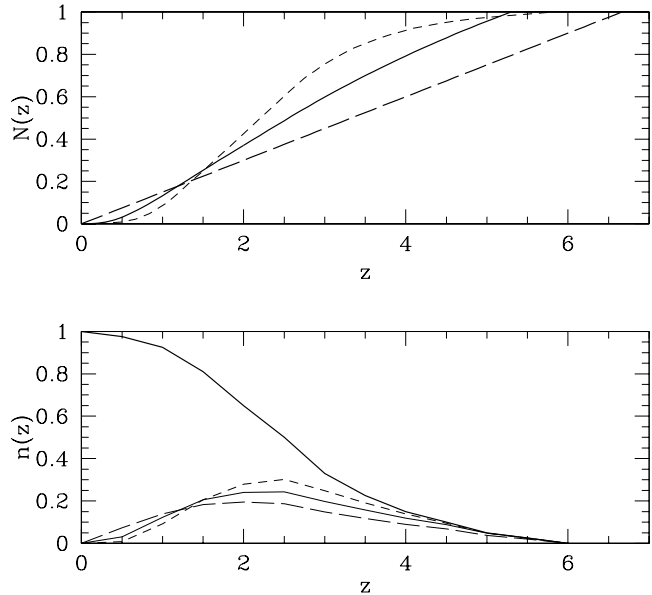}
\caption{Top panel: The predicted cumulative fraction of GRBs as a
  function of redshift for various models studied here. The solid
  curve is the prediction for Model II, the short-dashed curve for
  Model III and the long-dashed curve is for Model IV. Note that
  Models III and IV are normalized with respect to Model II.  Bottom
  panel: We illustrate the effect of folding in the detection
  efficiency of SWIFT as currently modeled by Mortsell \& Sollerman
  (2005). Despite the fall-off in sensitivity for $z > 6$ bursts as
  indicated by the thin solid line curve, the SFR models will still be
  distinguishable for about 1000 detected bursts. The conventions for
  the line-type are as before: thick solid [Model I]; long-dashed
  [Model IV] and short-dashed [Model III].}
\end{center}
\end{figure}

We then optimize to determine the value of the 3 free parameters in
the LF, $\gamma$ and $L_0$ and the normalization constant, for the
different SF history models considered here. The overall quality of
the best--fit in the $\chi$-square \footnote{The best-fit
$\chi^2$/d.o.f for the models ranges from 0.76 -- 1.05.}sense is
slightly better for our semi-analytic SF model (Model II) as it is
slightly lower at high redshifts ($z > 5$) compared to the Porciani \&
Madau model. This is due to the fact that increasing star formation at
high redshift causes the over-prediction of the number of bursts to be
consistent with the faintest off--line BATSE counts. We find that
increasing the steepness of the high--luminosity tail of $\psi(L)$
requires an increase in $L_0$ for the same value of $\chi^2$, implying
that both models studied here need the presence of relatively
high--luminosity events to reproduce the data.  On comparing the
properties of the LFs that provide the best--fit for each star
formation rate, we also find that the typical burst luminosity
increases in models with larger amounts of star formation at early
epochs as also shown by Lloyd-Ronning et al. (2002).  However, two of
the star formation models considered here, Model I and Model II,
predict a very similar redshift distribution for GRBs. This is due to
the fact that although the SF for these models appears to diverge at
$z > 7$, there is not much time elapsed at these high
redshifts. Therefore, we only show the predicted distribution for
Model II in Fig.~2. For our Model III, with significantly lower
amounts of star formation, fewer bursts are predicted at higher
redshift compared to Models I and II. Folding in the detection
efficiency model curve for SWIFT (Mortsell \& Sollerman 2005), we
predict the observed GRB redshift distribution for SWIFT (see Fig.~2).

\section{Results and Conclusions}

Our best-fit parameters for the LF of GRBs, combined with BATSE number
counts and the peak-flux distribution for observed bursts are then
used to predict the redshift distribution for GRBs given a SFR
model. The results for the SFR models and the metallicity dependent
rate model are plotted in Fig.~2. The predicted z-distribution for
Models I and II are very similar, and we show the curve only for Model
II. For both Models I and II, we find that a very large fraction
$\sim$ 75\% of all GRBs originate at redshifts of four or lower. Our
results are in excellent agreement with those reported recently by
Guetta et al. (2005), where they studied the luminosity and angular
distribution of long duration GRBs, similarly modeling the SFR history
(in particular see their Fig.~3). Note that Guetta et al. utilize the
relation between an assumed jet angular distribution and the GRB LF to
predict the observed redshift distribution of bursts. They predict a
local GRB detection rate for both the structured jet model and the
universal jet model that is corrected for beaming. In this work we
assume that the energy release in GRBs is isotropic.  In an earlier
calculation Bromm \& Loeb (2002) predicted a higher proportion of GRBs
at higher redshifts compared to this work.  This discrepancy arises
due to the fact the their treatment did not include a LF for GRBs and
did not take into account the spectral energy distribution of
GRBs. Unlike supernovae, GRBs are not standard candles, although there
has been a recent claim of a tight correlation between the rest-frame
peak energy and the rest-frame beaming corrected gamma-ray energy
release (Ghirlanda et al. 2004) which may allow them to be used as
standard yard-sticks for cosmography purposes (Mortsell \& Sollerman
2005). The inclusion of the LF coupled with the Band function are key
ingredients that are needed in order to make robust predictions for
the redshift distribution, even though there are considerable
uncertainties. For Model III with lower SFRs at high redshifts, we
find a much smaller fraction of GRBs only about $\sim\,10\%$ to
originate from $z > 4$. It is interesting to note that our toy Model
IV predicts (not surprisingly) a higher proportion of bursts $\sim
40\%$ at $z > 4$.

Our Model IV assumes that as low-mass galaxies are likely to have
statistically lower metallicities, they are likely to contain more
luminous GRBs than high-mass galaxies. Given that galaxies 
assemble hierarchically through mergers, then it is also possible
that the highest redshift GRBs could be systematically more luminous
owing to the lower mass and metallicity of their hosts. Such an effect
motivates the metallicity dependence of the GRB rate assumed
here. Additionally, SF activity is likely to be enhanced in merging
galaxies. In major mergers of gas-rich spiral galaxies, this
enhancement takes place primarily in the inner kiloparsec. Metallicity
gradients in the gas are likely to be smoothed out, by both mixing
prior to SF and supernova enrichment during the star burst. GRB
luminosities could thus be suppressed in such well-mixed galaxies,
making GRBs more difficult to detect in these most luminous objects,
in which a significant fraction of all high-redshift star formation is
likely to have occurred. Shocks in tidal tails associated with merging
galaxies are also likely to precipitate the formation of high-mass
stars, yet as tidal tails are likely to consist of relatively
low-metallicity gas, it is perhaps these less intense sites of star
formation at large distances from galactic radii that are more likely
to yield detectable GRBs. As more SWIFT bursts are followed-up and
their environments are better studied, this correlation will be
testable.

Given the current uncertainties and our lack of knowledge of high
redshift star formation, if SWIFT detects a handful of bursts from
beyond $z \sim 6$ with measured redshifts, these bursts might end up
providing the only observational constraint on the SF at these early
epochs (see Fig.~2). We find that for a large number, of say, 1000
detected bursts, we will be able to discriminate between the various
SFR models as illustrated in Fig.~ 3. Note that a more accurate and
calibrated detection efficiency curve for SWIFT will be available
after a few years of operation. The robustness of the assumption that
the SFR is a good proxy for the GRB rate can also be tested further in
the near future as the uncertainties due to dust correction in
determining SFRs are better understood and the host galaxies of GRBs
are studied in more detail. As we explore here, the relation between
averaged metallicity of the Universe and GRB rate might also prove to
be testable with future observations of GRB host galaxies.

\section*{Acknowledgments} 

We thank Elena Rossi, Daphne Guetta, Eli Waxman and an anonymous referee 
for useful suggestions that have improved the paper.

\end{document}